%% file: main.tex
\documentclass[prb,twocolumn,amsmath,amssymb,floatfix,superscriptaddress]{revtex4-2} 
\usepackage{graphicx}
\usepackage{epstopdf}  
\usepackage{bm}
\usepackage{amsmath,amssymb}
\usepackage{dcolumn}
\usepackage{bm}
\usepackage{epsfig}
\usepackage{longtable}
\usepackage{subfigure}
\usepackage{multirow}
\usepackage{color}

\input{newcommand}

\DeclareUnicodeCharacter{2212}{-}
\begin{document}

\title{{\rm\small\hfill }\\
Developing correlation-consistent numeric atom-centered orbital basis sets for Krypton: Applications in RPA-based correlated calculations}

\author{Sixian Yang}
\affiliation{Key Laboratory of Quantum Information, University of Science and Technology of China, Hefei, 230026, China}
\affiliation{Institute of Physics, Chinese Academy of Sciences, Beijing 100190, China}
\author{Igor Ying Zhang}
\email{igor\_zhangying@fudan.edu.cn} 
\affiliation{Collaborative Innovation Center of Chemistry for Energy Materials, Shanghai, 
 Key Laboratory of Molecular Catalysis and Innovative Materials,
 MOE Key Laboratory of Computational Physical Sciences, 
 Shanghai Key Laboratory of Bioactive Small Molecules,
 Department of Chemistry, Fudan University, Shanghai 200433, China}
\author{Xinguo Ren}
\email{renxg@iphy.ac.cn}
\affiliation{Institute of Physics, Chinese Academy of Sciences, Beijing 100190, China}


\begin{abstract}
Localized atomic orbitals are the preferred basis-set choice for large-scale explicit correlated calculations, and high-quality hierarchical correlation-consistent basis sets are a prerequisite for correlated methods to deliver numerically reliable results.
At present, Numeric Atom-centered Orbital (NAO) basis sets with valence correlation consistency (VCC), designated as NAO-VCC-$n$Z, are only available for light elements from hydrogen (H) to argon (Ar) (\textit{New J. Phys.} \textbf{15}, 123033, (2013) ). In this work, we extend this series by developing NAO-VCC-$n$Z basis sets for krypton (Kr), a prototypical element in the fourth row of periodic table. We demonstrate that NAO-VCC-$n$Z basis sets facilitate the convergence of electronic total-energy calculations using the Random Phase Approximation (RPA), which can be used together with a two-point extrapolation scheme to approach the complete-basis-set (CBS) limit. Notably, the Basis Set Superposition Error (BSSE) associated with the newly generated NAO basis sets is minimal, making them suitable for applications where BSSE correction is either cumbersome or impractical to do. After confirming the reliability of NAO basis sets for Kr, we proceed to calculate the Helmholtz free energy for Kr crystal at the theoretical level of RPA plus renormalized single excitation (rSE) correction. From this, we derive the pressure-volume ($P$-$V$) diagram, which shows excellent agreement with the latest experimental data. Our work demonstrates the capability of correlation-consistent NAO basis sets for heavy elements, paving the way toward numerically reliable correlated calculations for bulk materials.
\end{abstract}

\maketitle
\section{Introduction}

 Kohn-Sham (KS) density functional theory (DFT) \cite{Hohenberg/Kohn:1964,Kohn/Sham:1965} has become an indispensable tool for computational studies 
 in physics, chemistry, and materials science. Practical applications of DFT rely on useful approximations to the exchange-correlation (XC) 
 functional -- the only unknown component in KS-DFT. The success of DFT lies largely in the fact that good compromise between accuracy
 and the computational cost can be achieved at the level
 of local and semi-local approximations \cite{Kohn/Sham:1965,Perdew/Zunger:1981,Perdew/Burke/Ernzerhof:1996,Sun/Ruzsinszky/Perdew:2015}. Despite their success, these functionals suffer from intrinsic deficiencies, such
 as severe delocalization errors, the lack of long-range van der Waals interaction, and the incapability of dealing with strong electron
 correlations, to name a few \cite{Cohen/Sanchez/Yang:2012}. To further improve the accuracy of first-principles approaches, recent advancements have focused on the development of so-called ``fifth-rung'' functionals. These include the Random Phase Approximation (RPA) \cite{Bohm/Pines:1953,Eshuis/Bates/Furche:2012,Ren/etal:2012b,Ren/etal:2012,Ren/etal:2013} formulated within
 the adiabatic-connection fluctuation-dissipation theorem framework \cite{Langreth/Perdew:1977,Gunnarsson/Lundqvist:1976}, and
 doubly hybrid functionals \cite{Zhao/Lynch/Truhlar:2004,Grimme:2006,Zhang/Xu/Goddard:2009,Zhang/Xu:2014} that integrate Density Functional Approximations (DFAs) with explicit non-local correlated methods, such as
the M{\o}ller–Plesset perturbation theory (MP2) \cite{Moller/Plesset:1934}.
One significant challenge associated with these fifth-rung functionals (and explicit correlated methods in general)
is the need for a large amount of unoccupied single-particle orbitals. In practice, the corresponding correlation energy 
exhibits slow convergence with respect to the basis set size \cite{Lebegue/etal:2010,Harl/Schimka/Kresse:2010,Ren/etal:2012b,Eshuis/Furche:2012,Janesko/Henderson/Scuseria:2009}. This issue stems from the inadequate representation of the sharp Coulomb cusp at the electron-electron coordinate coalescence, where the first derivative of the 
many-electron wave function is discontinuous. Achieving an accurate description of such a cusp using smooth orbital expansions necessitates a large number of basis functions \cite{tew2007electron,noga1992cc}. Over the years, considerable efforts have been devoted to addressing the slow convergence of the electron correlation energy. One prevalent approach is the Complete Basis Set (CBS) extrapolation scheme \cite{Grueneis/etal:2009,Harl/Schimka/Kresse:2010,Eshuis/Furche:2012,Furche:2001,Fabinao/DellaSala:2012,Grueneis/Marsman/Kresse:2010}, which 
is widely adopted by the quantum chemistry community to mitigate the Basis Set Incompleteness Errors (BSIE). This is accomplished by utilizing hierarchical basis sets allowing the convergence pattern of the correlation energy to be described through straightforward analytical formula. Another approach involves the utilization of explicit wave-function methods, such as the $F_{12}$ method, \cite{noga1992cc,hattig2012explicitly,kong2012explicitly,kutzelnigg1991wave}, which explicitly account for the electronic cusp and thereby reduce BSIE at finite basis sets.

Both techniques mentioned above have proven to be successful for light elements and molecular systems. Gaussian-Type Orbitals (GTOs) are often the basis set
choice for molecular calculations due to their nice analytical properties that facilitate the computation of multi-center integrals. In this regard,
the correlation-consistent GTOs developed by Dunning \cite{Dunning:1989}, known as ``cc-pVXZ'' basis sets, have gained widespread acceptance in the field of quantum chemistry. Their systematic convergence behavior allows for reliable extrapolation to the CBS limit. In recent years, there has been considerable interest
in applying quantum chemistry methods and fifth-rung functionals to extended periodic systems \cite{wang_doubly_2021,zhang_coupled_2019}. In such cases, the standard cc-pVXZ basis sets, while effective for molecular calculations, are not optimal. 
One issue is that the extended tails of GTOs can result in
an overlap matrix with a large condition number for extended periodic systems, rendering the preceding self-consistent calculations unstable
before a sufficiently large GTO basis set can be employed for a reliable extrapolation. 
Secondly, the Basis Set Superposition Error (BSSE) associated with commonly used cc-pVXZ basis sets can be significant for correlation methods, which is
difficult to correct in periodic systems. Within the quantum chemistry community, there is ongoing research aimed at developing GTOs better suited
for correlated calculations in periodic systems. For example, Ye ~\textit{et al.} \cite{ye2022correlation} recently developed the ``GTH-cc-pVXZ'' 
GTOs, which are combined with the Goedecker−Teter−Hutter (GTH) type pseudopotentials  \cite{Goedecker/Teter/Hutter:1996,Hartwigsen/Goedecker/Hutter:1998}, to facilitate solid-state calculations using MP2 and Coupled Cluster theory at the level of Single, Double, and perturbative Triple excitations (CCSD(T)).

Alternatively, plane-wave basis sets can be used for correlated calculations in solid-state systems \cite{Marsman/etal:2009,Sphepherd/etal:2012}. However, it should be noted that the pseudopotentials designed for
ground-state DFAs may not be suitable for such correlated calculations. For example, special pseudopotentials must be developed for 
high-precision $GW$ calculations in the projector augmented wave framework \cite{Klimes/etal:2015}. Furthermore, plane-wave basis sets make it challenging
to formulate low-scaling algorithms that fully exploit Kohn's nearsightness principle \cite{Kohn:1996}. Although $O(N^3)$-scaling algorithm
has been developed for RPA and $GW$ utilizing the fast Fourier transform between real and reciprocal space \cite{Rojas/Godby/Needs:1995} and the efficient Fourier transform between the imaginary time and frequency domains \cite{Klimes/Kaltak/Kresse:2014}, further scaling reductions that make use of the intrinsic locality of the electronic structure have not yet been reported. 

In the case of Numeric Atom-centered Orbitals (NAOs), the situation is even less satisfactory. In 2013, Zhang~\textit{et al.} developed the valence-correlation-consistent NAO basis sets, termed as NAO-VCC-$n$Z \cite{IgorZhang/etal:2013}, which are apt for correlated calculations in molecular calculations. However, NAO-VCC-$n$Z is currently only available for light elements from H to Ar. For solid-state calculations, these basis sets require further refinement and modification to alleviate the ill-conditioning problem \cite{IgorZhang/etal:2019}, which is more pronounced in solids 
compared to molecules when using large basis sets. Consequently, there is an urgent need to develop correlation-consistent NAO basis sets that are
1) applicable to heavier elements beyond Ar, and 2) suitable for solid-state calculations.

In this work, we first construct hierarchical valence-correlation-consistent NAO basis sets for the element krypton (Kr), largely adhering to the prescription
developed in Ref.~\cite{IgorZhang/etal:2013}. Specifically, the basis functions are generated by minimizing the frozen-core PBE-based RPA total energy
for a single atom, while subject to the ``even-tempered'' principle, akin to the approach
used in the construction of cc-pVXZ basis sets. 
The candidate basis functions are all ``hydrogen-like'' functions \cite{Blum/etal:2009}.
We then assess the performance of the newly generated NAO-VCC-$n$Z ($n=$2-5) basis sets for both Kr dimer (Kr$_2$) and Kr crystal systems. Systematic convergence behaviors
of the binding energies are observed for both the dimer and bulk systems, allowing for reliable extrapolation to the CBS limit. More importantly, our comparison with and without counterpoise corrections reveals that these new NAO basis sets carry rather small BSSE, permitting reliable calculations even without BSSE corrections. 
In contrast, the cc-pVXZ GTOs and FHI-aims-2009 (\textit{tier}'s) NAOs for Kr are contaminated with huge BSSE, resulting in the binding energies that necessitate BSSE corrections for obtaining meaningful results.
The minimal and controllable BSSE in the newly generated NAO-VCC-$n$Z for Kr offers a significant advantage, especially in situations 
where BSSE corrections are cumbersome or impractical to perform.
Finally, we calculate the Helmholtz free energy of the  face-centered-cubic (FCC) Kr crystal, based on which the $P$-$V$ curve at finite temperatures can be attained. The electronic part of the free energy is obtained from converged RPA+rSE calculations. The final RPA+rSE based $P$-$V$ curve shows excellent agreement with experimental measurements for compressed Kr crystal across a broad pressure range. The experience gained from developing these
correlation-consistent NAO basis sets for Kr is highly useful for guiding similar efforts for other heavy elements. Works on constructing correlation-consistent NAO basis sets for other fourth-row elements are still ongoing, and will be published elsewhere.

The paper is organized as follows. Sec.~\ref{sec:methods} outlines the foundamentals of NAO basis sets, focusing on the prescription used to
generate correlation-consistent NAOs for Kr, along with the computational details of the present work. The major results are presented in
Sec.~\ref{sec:result}, including the detailed layout of the NAO-VCC-$n$Z basis sets for Kr (Sec.~\ref{sec:results:NAO}), a systematic study of the convergence behavior of the new basis sets for both Kr$_2$ and the bulk system (Sec.~\ref{sec:results:binding}), and the $P$-$V$ curve based on converged RPA+rSE calculations (Sec.~\ref{sec:results:PV}). The paper concludes in Sec.~\ref{sec:conc}.

\section{\label{sec:methods}Methodology and computational details}
\subsection{Numerical atomic-centered basis sets}
In this work, all calculations are performed using the \textit{all-electron} Fritz Haber Institute \textit{ab initio molecular simulations} (FHI-aims) code package \cite{Blum/etal:2009}. The primary basis-set choice of FHI-aims is NAO $\{\phi_i (\bfr)\}$, given by
numerically tabulated radial functions multiplied by real spherical harmonics,
\begin{equation}
  \phi_{i}(\bfr)=\frac{u_{i}(r)}{r}Y_{lm}(\Omega) \, ,
  \label{Eq:NAO orbital}
 \end{equation}
 where $Y_{lm}(\Omega)$ denotes real parts ($m=0,\cdots,l$) and imaginary parts ($m=-l,\cdots,-1$) of complex spherical harmonics.
 In FHI-aims, one has the flexibility to use radial functions with analytic forms, such as Slater-Type Orbitals (STOs) or GTOs.
 Generally, the radial part of the NAOs  $u_i(r)$ is obtained by solving the following radial Schr\"{o}dinger equation, 
\begin{equation}
  \left[-\frac{1}{2}\frac{d^{2}}{dr^{2}}+\frac{l(l+1)}{2r^{2}}+v_{i}(r)+v_{cut}(r)\right]u_{i}(r)=\epsilon_{i}u_{i}(r) \, .
  \label{Eq:radial function}
 \end{equation}
 In Eq.~\ref{Eq:radial function}, the total effective potential contains two parts: a radial potential $v_{i}(r)$, which governs the main shape of $u_{i}(r)$, and a confining potential $v_{cut}(r)$ which is switched on beyond a certain distance $r_{onset}$ from the nucleus and becomes infinite at $r_{onset}+w$. Careful tests indicate that $w=2.0$ \AA\ is a good default value, which ensures a smooth decay of the radial tails of
 $u_{i}(r)$ to strict zero at $r_{onset}+w$. 
 The choice of the cutoff radius $r_{onset}$ of the confining potential $v_{cut}(r)$ has been tested by Zhang~\textit{et al.} \cite{IgorZhang/etal:2013} and 
 it was shown that the energy variation is at the level of meV for $r_{onset}$ changes from 4.0 to 6.0 \AA. This uncertainty is negligible compared
 to the typical BSIE for correlation methods \cite{Ren/etal:2012b}.

In FHI-aims, NAO basis sets feature the use of so-called ``minimal basis functions'', which are generated from self-consistent, 
non-spin-polarized, and
spherically symmetric free-atom radial potential $v_{i}(r)=v_{at}^{free}(r)$. The minimal basis consists of core and valence functions for all occupied states, exhibiting excellent performance by naturally capturing wavefunction oscillations near the nucleus, where the nuclear $Z/r$ potential dominates. Hence the minimal basis serves as a natural foundation for constructing NAO basis sets for correlated methods. Also, it is worth noting that the description of the minimal basis of NAOs requires less dense radial grids for light elements compared to GTOs such as Dunning's cc-pVXZ basis, as demonstrated in
Ref.~\cite{zhang2013numeric}. 



\subsection{NAO-VCC-$n$Z basis sets with valence correlation consistency}
\label{sec:NAO-VCC-Kr}
Currently, there exists a set of optimized NAO basis sets, referred to as FHI-aims-2009 or more commonly as ``\textit{tier}-$n$''
basis sets \cite{Blum/etal:2009}, which are the standard choice for the ground-state calculations in FHI-aims. FHI-aims-2009 basis sets are 
constructed by a step-wise minimization of total energy from LDA calculations of selected dimers. Studies employing \textit{tier}-$n$ basis sets for conventional (semi)-local and hybrid DFAs have demonstrated their high efficacy \cite{Blum/etal:2009,Ren/etal:2012}. However, reliable results for correlation methods that require virtual states can only be achieved when a counterpoise (CP) correction is applied. In 2013, a new sequence of NAO basis sets, inspired by Dunning's correlation-consistency strategy, was developed \cite{zhang2013numeric}. These NAO basis sets, designated as "NAO-VCC-$n$Z", achieve systematic convergence of the frozen-core RPA total energy to the CBS limit. More importantly, a two-point extrapolation to the CBS limit effectively mitigates the BSSE issue, which is important in situations where a CP correction is impractical. Unfortunately, these NAO-VCC-$n$Z basis sets are currently limited to light elements from H to Ar.

In this work, we extend the work of Ref.~\cite{IgorZhang/etal:2013,IgorZhang/etal:2019} to construct the NAO-VCC-$n$Z basis sets for a prototypical fourth-row main-group element -- Kr. We group our NAO basis into  ``one base + two subsets'', where ``one base'' refers to the minimal basis described above, and the ``two subsets'' include: i) the ($sp$) correlation set, ii) the polarization set on top of the FHI-aims original minimal basis. Here, ($sp$) indicates that only $s$- and $p$-type orbitals are used, and the polarization set includes NAOs with higher momenta, i.e.~$d, f, g$ functions, etc. The $4s^{2}$ $4p^{6}$ valence-shell configuration is adopted for Kr in post-KS RPA calculations, while occupied $3d$ and lower states are frozen. We employ radial functions $u_{i}(r)$ generated via hydrogen-like radial potential $v_{i}(r)=z_{i}/r$ for all angular momenta $l$, and each radial function yields $2l+1$ basis functions by multiplying the real spherical harmonics (cf. Eq.~\ref{Eq:NAO orbital}). Here we set the principal quantum number $n=l+1$, resulting in nodeless radial functions similar to STOs \cite{van2003optimized,te2001chemistry}.

The effective nuclear charge $\{z_{i}\}$, which determines the actual shape of hydrogen-like basis functions, is the major parameter
to be optimized. For those nodeless radial functions, we use ``even-tempered"\cite{raffenetti1973even}
scheme to expand the effective nuclear charges:
\begin{equation}
   z_{i}=\alpha\beta^{i-1} \qquad  i=1,2,...,N_{orb}\, .
  \label{Eq:even-tempered}
 \end{equation}
 Here the nuclear charges $\{z_{i}\}$ are represented by two parameters $(\alpha,\beta)$ which are the actual parameters to be optimized. 
 Individual subsets of radial functions with the same angular momentum $l$ share the same $(\alpha,\beta)$ parameters. For example, in the case of a quadruple–$\zeta$ basis 
 one could expand all three $s$-type functions via the optimization of a pair of parameters $(\alpha_{1},\beta_{1})$, while a new pair of parameters $(\alpha_{2},\beta_{2})$ should be used for three $d$-type functions in the polarization set, and so on. 

 Following Ref.~\cite{zhang2013numeric}, we chose the frozen-core RPA@PBE energy of a single Kr atom as the optimization target for the parameters characterizing the basis functions. 
 Our optimization procedure proceeds as follows: Initially, the $sp$ correlation subset is optimized, followed by the optimization of the polarization subset. Subsequently, the $sp$ correlation subset undergoes a re-optimization as the final step. The Nealdar-Mead downhill simplex algorithm \cite{nelder1965simplex} is used for executing the optimization of the parameters $(\alpha,\beta)$, with a convergence criterion of $10^{-3}$ for the effective nuclear charge $Z$. This corresponds to an approximate convergence criterion of $10^{-6}$ eV for the RPA total energy.
To search for the global minimum, we explored various initial guesses for $(\alpha,\beta)$ within a reasonable window, using a tabulated grid.

\subsection{Computational details}
The total energy of RPA as a post-KS computational method is given by
\begin{equation}
   E^\text{RPA}=E^\text{DFA}-E_{xc}^\text{DFA}+E_{x}^\text{EX}+E_{c}^\text{RPA} \, ,
  \label{Eq:RPA compound}
 \end{equation}
whereby the XC component $E_{xc}^\text{DFA}$ of the total energy of a conventional DFA, $E^\text{DFA}$, is replaced by the exact-exchange energy $E_{x}^\text{EX}$ plus RPA correlation energy $E_{c}^\text{RPA}$, both evaluated in terms of pre-determined DFA orbitals and orbital energies. 
Furthermore, a ``single excitation” (SE) 
term, and in particular its renormalized version (termed as renormalized SE, or rSE) $E_{c}^\text{rSE}$ can be added to the standard RPA
total energy to alleviate the underbinding behavior of RPA for molecules and solids.  This correction scheme has been discussed in Refs.~\cite{Ren/etal:2011,Ren/etal:2013}, and subsequently
in Ref.~\cite{Klimes/etal:2015}. Benchmark tests for NAO-based RPA and RPA+rSE calculations have demonstrated the excellent performance of 
these computational schemes for accurately describing both molecular and periodic systems \cite{Ren/etal:2012,Ren/etal:2013,IgorZhang/etal:2019,yang2022phase,tahir2022localized}.

The employment of NAO-VCC-$n$Z basis sets allows the RPA or RPA+rSE total energy to be extrapolated
to the CBS limit via the following formula \cite{Halkier/etal:1998,Helgaker/etal:1997},
\begin{equation}
    E(n) = E(\infty) - C/n^3\, .
    \label{eq:basis_dependence}
\end{equation}
In Eq.~\ref{eq:basis_dependence}, $n$ is the cardinal number of the NAO-VCC-$n$Z basis sets, and $C$ is a fitting parameter.
Equation~\ref{eq:basis_dependence} is a widely used two-point extrapolation formula initially developed for correlation-consistent Gaussian basis sets \cite{Helgaker/etal:1997},
but also works well for NAO-VCC-$n$Z basis sets \cite{IgorZhang/etal:2013}.  In this work, basis sets with $n$=4, 5 are used for
the extrapolation, implying that 
 \begin{equation}
  E(\infty)=\frac{E(4)4^{3}-E(5)5^{3}}{4^{3}-5^{3} } \, .
  \label{Eq:extrapolate}
 \end{equation}
 The CBS limit results obtained via Eq.~\ref{Eq:extrapolate} are denoted as CBS(4,5).
Conventional DFAs that require only occupied states converge much faster with respect to  the basis set size. 
In this work, a finite NAO-VCC-5Z basis set
is used for conventional DFA calculations, which proves to be adequate. Also, for periodic systems, a finite $\bfk$-point mesh is used for Brillouin-zone sampling. For FCC primitive cell, an $8\times 8 \times 8$ $\bfk$ grid is used , which is sufficient to achieve sub-meV precision for the 
RPA total energy. Moreover, the scaled zeroth-order regular approximation (ZORA) \cite{Lenthe/Baerends/Snijders:1994} scheme is employed to account
for the relativistic effect.

For free energy calculations, one needs to take into account of the entropy contribution from lattice vibrations.  
In this work, we calculate the phonon spectra at the PBE level, and the Helmholtz free energy at a given volume $V$ and temperature $T$ invoked to determine the $P$-$V$ diagram is given as
\begin{align}
  F(V,T)=& E^\text{RPA+rSE}(V)+ \frac{1}{2}\sum_{\bfq,j}\hbar\omega_{\bfq,j}(V) + \nonumber \\
         & \emph{k}_{B}T\sum_{\bfq,j}ln\left[2 sinh \left( \frac{\hbar\omega_{\bfq,j}(V)}{2\emph{k}_{B}T}\right) \right]
  \label{Eq:QHA}
 \end{align}
 Here the summation over $\bfq$ goes over the Brillouin zone and $j$ labels different phonon modes.
 Furthermore, in Eq.~\ref{Eq:QHA} $E^\text{RPA+rSE}(V)$ is the RPA+rSE total energy at 0 K, and the second and third terms correspond to the contributions from zero-point energies and lattice vibrations at finite temperatures. The phonon frequencies are calculated by the finite displacement method as implemented in the
 PHONOPY code \cite{phonopy} interfaced with FHI-aims. Convergence criteria for the total energy and force are set to $10^{-6}$ eV and $10^{-5}$ eV/A, respectively. A displacement of 0.002 A for the finite displacement method is adopted. In Ref.~\cite{yang2022phase}, the influence of the 
 employed functionals on the phonon spectrum and the subsequent free energy calculations was investigated, the conclusion was that the use 
 of PBE functional instead of the RPA+rSE in the phonon calculations should be adequate for calculating the $P$-$V$ curve.
 


\section{\label{sec:result}Results and discussions}
\subsection{\label{sec:results:NAO}NAO-VCC-$n$Z basis sets of Kr}

As discussed in Sec.~\ref{sec:NAO-VCC-Kr},  each of the NAO-VCC-$n$Z basis sets for Kr consists of a subset of $sp$ correlation functions, a subset of
polarization functions, and a base of minimal basis functions. The former two subsets are optimized in this work, and the layouts of these
two subsets for $n=2,3,4,5$ are shown in Table~\ref{Tab:form of NAO}. In addition, the same minimal basis functions of (4$s$ 3$p$ 1$d$) for Kr
are included in all NAO-VCC basis sets and added up in total count in  Table~\ref{Tab:form of NAO}. Note that the minimal basis functions
are determined using Eq.~\ref{Eq:radial function} and hence their exact shape will (slightly) depend on the actual XC potential used in
the SCF calculations.
\begin{table*}
\caption{Number of radial functions in Kr NAO-VCC-$n$Z ($n$=2,3,4,5) basis sets. The minimal basis set of (4$s$ 3$p$ 1$d$) is not shown in this table, but is added up in the total count.}
\begin{tabular}{@{\extracolsep{\fill}}lccccccc}
\hline\\[-1.5ex]
\qquad & (sp) correlation subset& & polarization subset & & total\\[0.3ex]
\hline \\[-0.3ex]
NAO-VCC-2Z   & (1$s$ 1$p$)     & & (1$d$)       & & (5$s$ 4$p$ 2$d$)\\
NAO-VCC-3Z   & (2$s$ 2$p$)     & & (2$d$ 1$f$)    & & (6$s$ 5$p$ 3$d$ 1$f$)\\
NAO-VCC-4Z   & (3$s$ 3$p$)     & & (3$d$ 2$f$ 1$g$) & & (7$s$ 6$p$ 4$d$ 2$f$ 1$g$)\\
NAO-VCC-5Z   & (4$s$ 4$p$)     & & (4$d$ 3$f$ 2$g$ 1$h$) & & (8$s$ 7$p$ 5$d$ 3$f$ 2$g$ 1$h$)\\
\hline \\[-1.5ex]

\end{tabular}
\label{Tab:form of NAO}
\end{table*}

\begin{figure}[t]
\centering
\includegraphics[width=0.45\textwidth,clip]{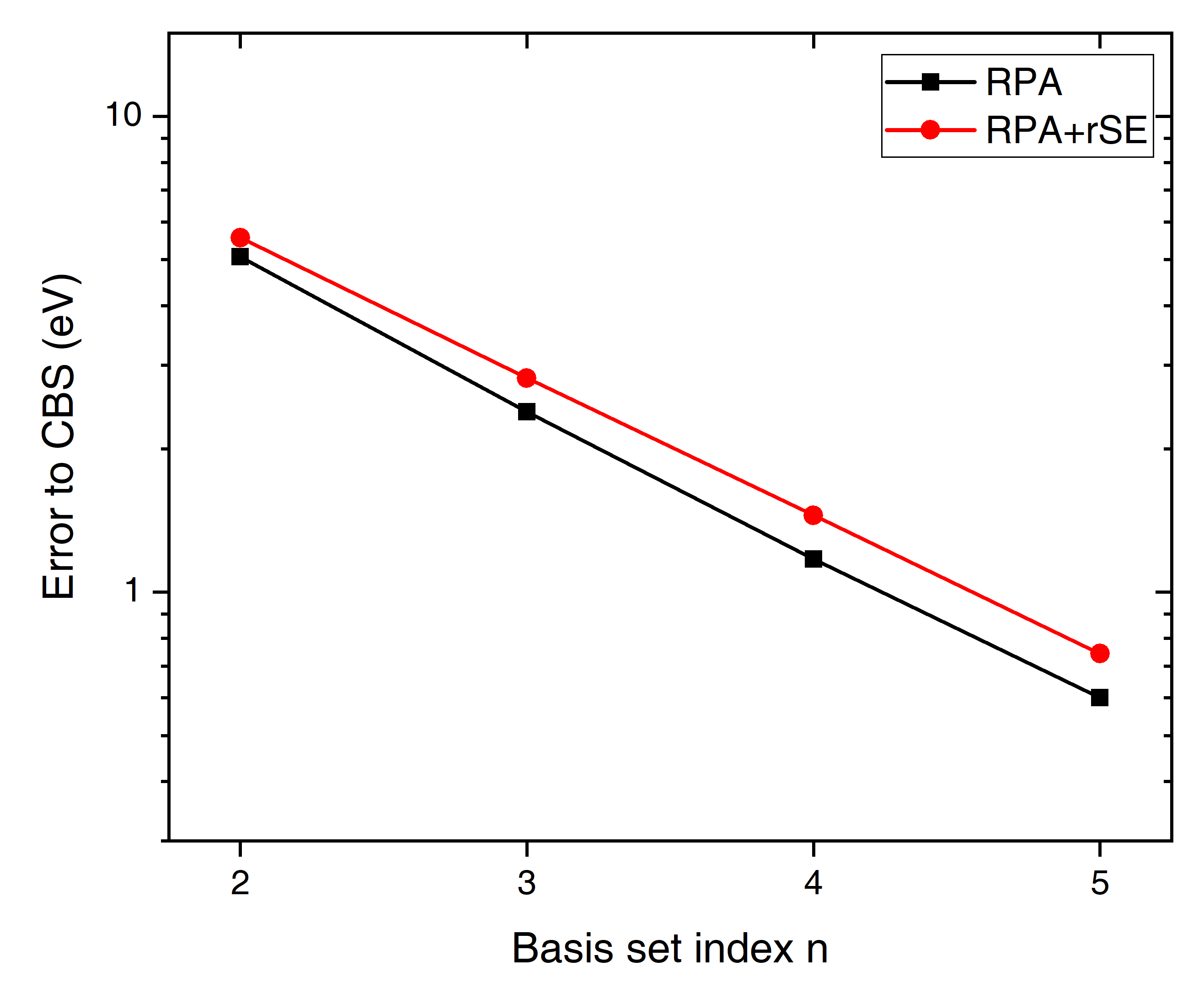}
\caption{\label{Fig:exponential convergence} Basis set errors of the RPA@PBE and (RPA+rSE)@PBE correlation energies for isolated Kr atom. 
The CBS(4,5) results are taken as the references. }
\end{figure}

Similar to the Dunning's cc-pVXZ basis sets \cite{Dunning:1989} and the NAO-VCC-$n$ basis sets for light elements \cite{zhang2013numeric},
the newly generated NAO-VCC-$n$Z basis sets for Kr adhere to the ``correlation consistency'' principle. Specifically, the increase of the correlation energy resulting from the inclusion of the $n$-th polarization function with angular momentum $l$ is approximately equal to the increase stemming from the inclusion of the $(n-1)$-th polarization function with angular momentum $l+1$. 
As illustrated in Fig.~\ref{Fig:exponential convergence}, the RPA and RPA+rSE correlation energies for the Kr atom exhibit a clear exponential convergence as a function of the basis set index $n$.
In Table. S1 of the Supplementary Materials (SM), the parameter settings of the NAO-VCC-$n$Z basis sets for Kr as used in FHI-aims calculations are presented.

\begin{figure*}
\subfigure{
\includegraphics[width=0.45\textwidth]{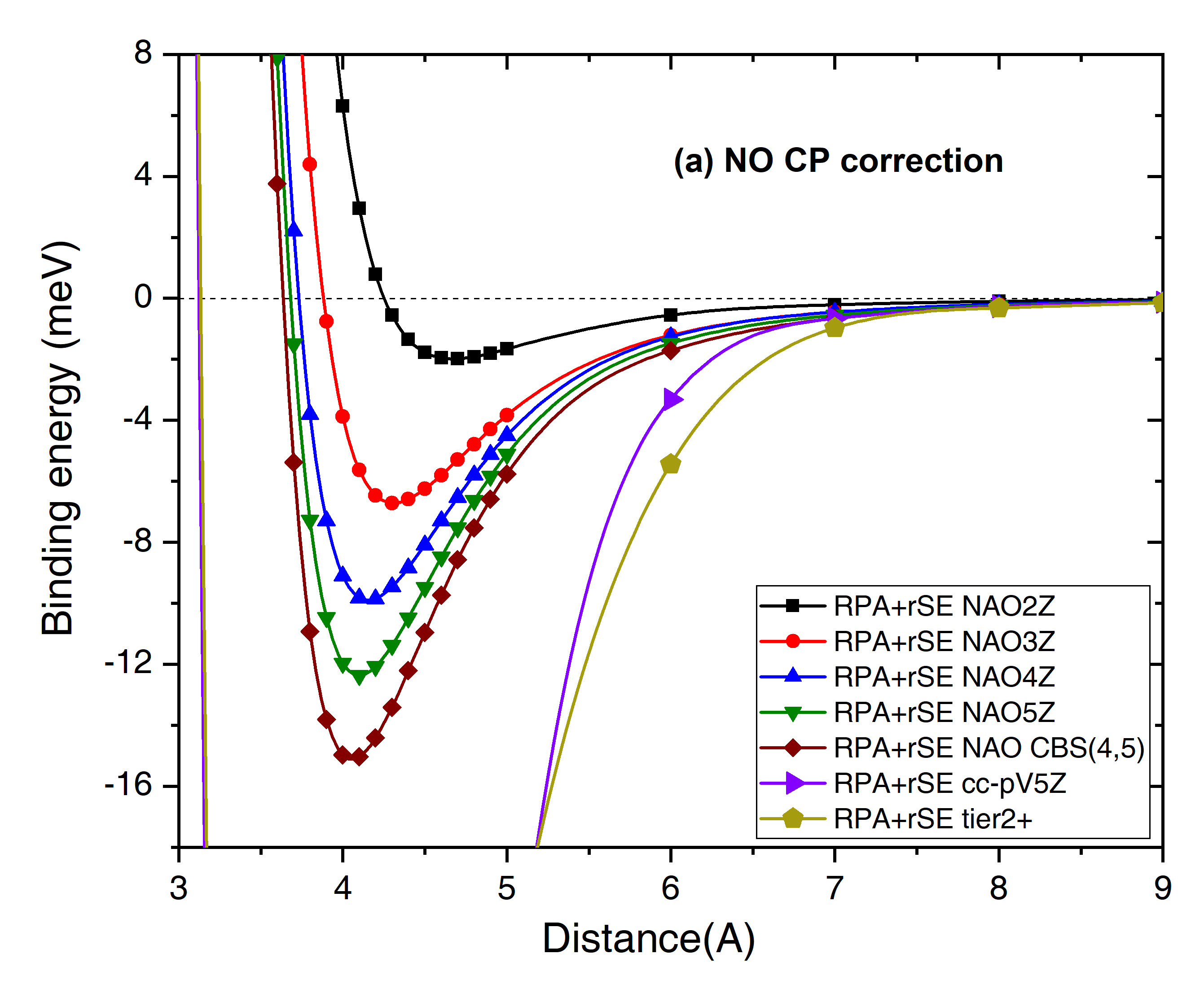}
}
\subfigure{
\includegraphics[width=0.45\textwidth]{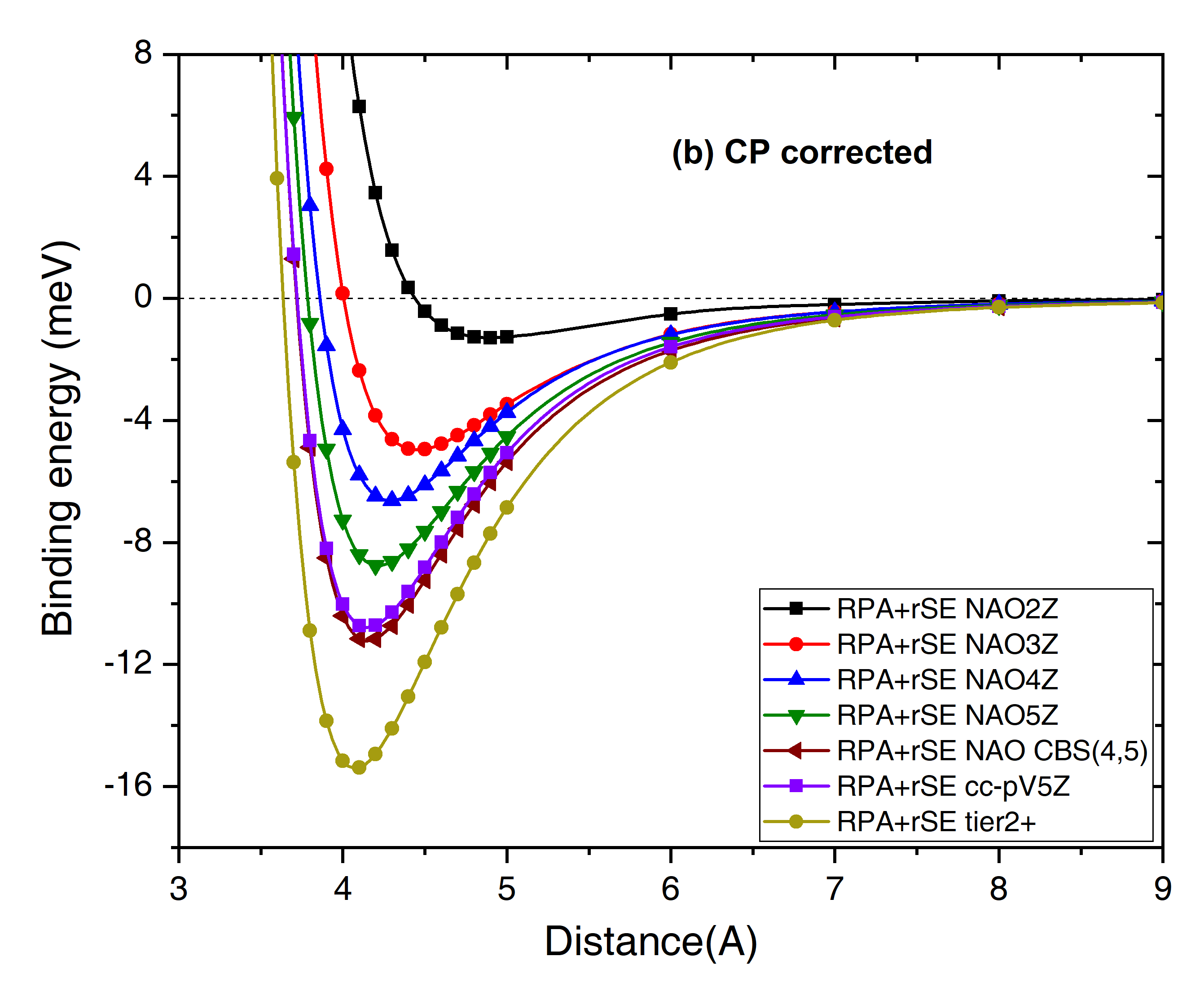}
}
\caption{\label{Fig:dimer} Binding energy curves of Kr$_2$ without CP correction(left panel) and with CP correction(right panel) are shown. NAO-VCC-$n$Z basis set is extrapolated to CBS limit CBS(4,5) and the other basis sets are using the largest size of its family (e.g. \textit{tier}2+, cc-pV5Z).}
\end{figure*}


\subsection{\label{sec:results:binding}Binding energy curves of the Kr dimer and solid bulk}

We now examine the performance of NAO-VCC-$n$Z basis sets on calculating the binding energies of the Kr dimer -- Kr$_2$. Figure~\ref{Fig:dimer} presents the (RPA+rSE)@PBE binding energy curves for Kr$_2$, which are calculated using NAO-VCC-$n$Z with $n=2$ to 5, as well as the extrapolated CBS(4,5) limit. The left and right panels of Fig.~\ref{Fig:dimer} show the results without and with CP corrections, respectively. 
Additionally, we performed RPA+rSE calculations using Gaussian cc-pV5Z and FHI-aims-2009 NAO basis sets for comparative analysis. For the latter, we choose \textit{tier} 2 plus some additional functions (denoted here as ``\textit{tier}2+''). It is the largest basis set of this type available for Kr. To isolate the impact of basis-set errors, a very tight convergence criterion is applied to numerical integration in both radial and angular parts. The calculated equilibrium bond lengths and the
corresponding binding energies for various basis sets are listed in Table~\ref{Tab:eq point of dimer}.

\begin{table*}
\caption{Equilibrium bond lengths $r_e$ (in \AA) and interaction energies $D_e$ (in meV) of RPA+rSE obtained using various different basis sets.  
Results obtained both with
(right side) and without (left side) CP corrections are presented.}
\begin{tabular}{@{\extracolsep{\fill}}lccccccc}
\hline\\[-1.5ex]
\multirow{2}{*}{Basis set} & \multicolumn{2}{c}{no CP correction}  & &
 \multicolumn{2}{c}{CP corrected}  \\[0.3ex]
\cline{2-3}  \cline{5-6}\\[-1.0ex]
& ~~~~~$r_e$~~~~~ & ~~~~~~$D_e$~~~~~ & ~~~ & ~~~~~$r_e$~~~~~ & ~~~~~~$D_e$~~~~~   \\[0.5ex]
\hline \\[-0.3ex]
NAO-VCC-3Z         & 4.31 &-6.71         & & 4.46 &-4.96    \\
NAO-VCC-4Z         & 4.15 &-9.90         & & 4.29 &-6.63    \\
NAO-VCC-5Z         & 4.10 &-12.37        & & 4.21 &-8.76    \\
NAO-VCC CBS(4,5)   & 4.05 &-15.12        & & 4.15 &-11.22   \\
\textit{tier}2+    & 3.58 &-90.92        & & 4.08 &-15.39   \\
cc-pV5Z            & 3.69 &-121.53       & & 4.15 &-10.79   \\

\hline \\[-1.5ex]

\end{tabular}
\label{Tab:eq point of dimer}
\end{table*}
Figure~\ref{Fig:dimer} shows that the newly generated NAO-VCC-$n$Z basis sets for Kr yield systematically converging RPA+rSE binding energies for the Kr$_2$. More importantly, a comparison between the left and right panels of Fig.~\ref{Fig:dimer} reveals that the BSSEs associated with NAO-VCC-$n$Z are rather small, corresponding to a small fraction of the total binding energy. In stark contrast, binding energies calculated using cc-pV5Z and the NAO \textit{tier}2+ basis sets are contaminated with substantial BSSE. Without performing CP corrections, both cc-pV$5$Z and \textit{tier}2+ significantly overestimate the binding energies by an order of magnitude, rendering the results entirely unusable. These discrepancies are evident in both Fig.~\ref{Fig:dimer} and Table~\ref{Tab:eq point of dimer}. Further results for other cc-pV$X$Z basis sets are presented in Fig.~S1 of the SM.

It should be noted that, by introducing an augmented set of diffuse functions, aug-cc-pV$X$Z basis sets usually outperform cc-pV$X$Z, in particular for weak interactions \cite{zhang2013numeric}. However, unlike aug-cc-pV$X$Z basis sets, our NAO-VCC-$n$Z basis sets do not incorporate an augmented set of diffuse functions. As shown in Fig.~S2 in the SM, the introduction of diffuse functions to NAO-VCC-$5$Z leads to a CP-corrected bonding energy of -13.70 meV, which is similar to the NAO-VCC-5Z result without CP corrections (-12.37 meV, see Table~\ref{Tab:eq point of dimer}).  This is appealing since in many situations it is too cumbersome or impractical to perform BSSE corrections, and in such situations the newly generated NAO-VCC-$n$Z basis set for Kr can be directly used, expecting that the small and controllable BSSE of this basis set compensates the effect of missing diffuse functions.

\begin{figure*}
\subfigure{
\includegraphics[width=0.45\textwidth]{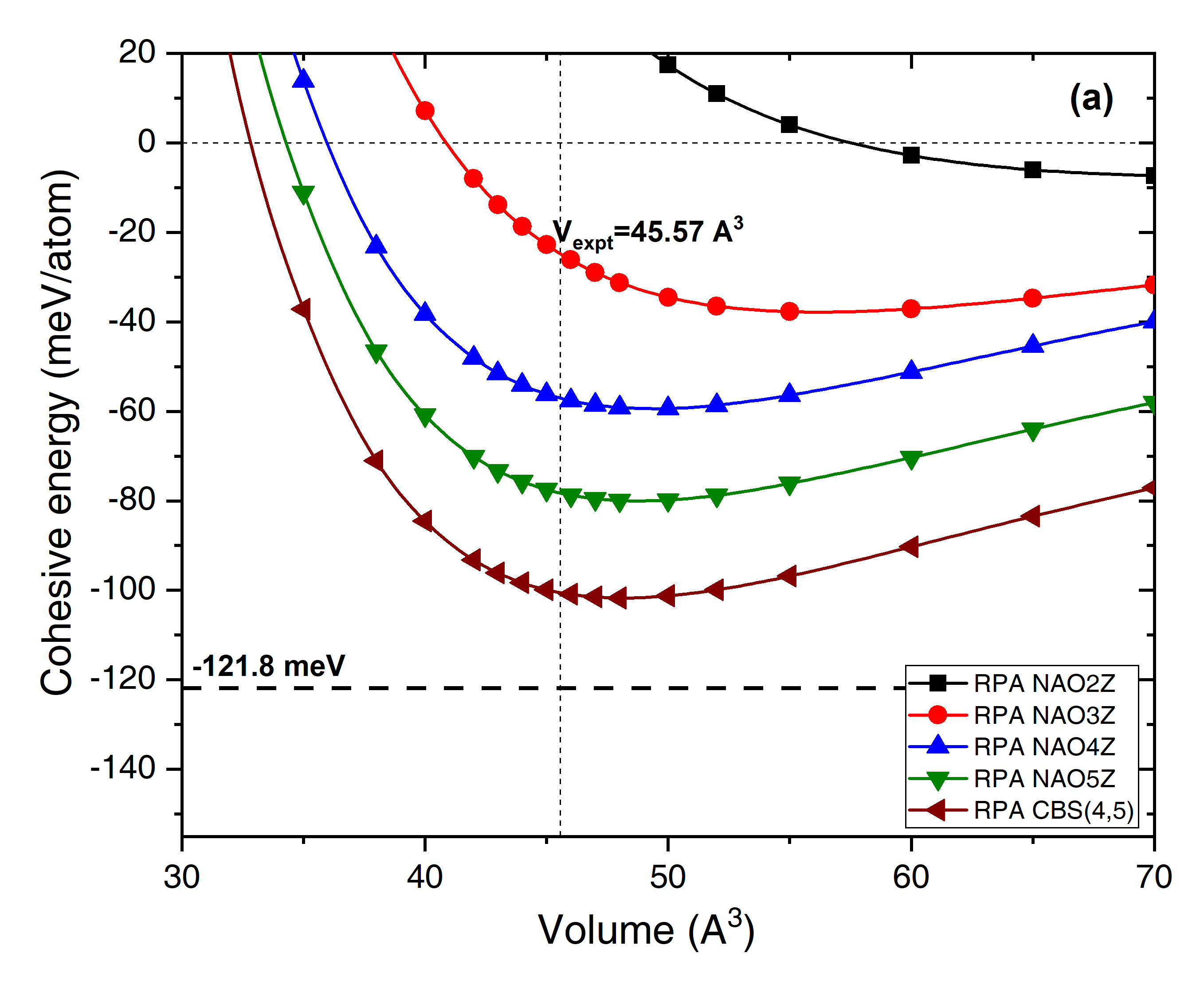}
}
\subfigure{
\includegraphics[width=0.45\textwidth]{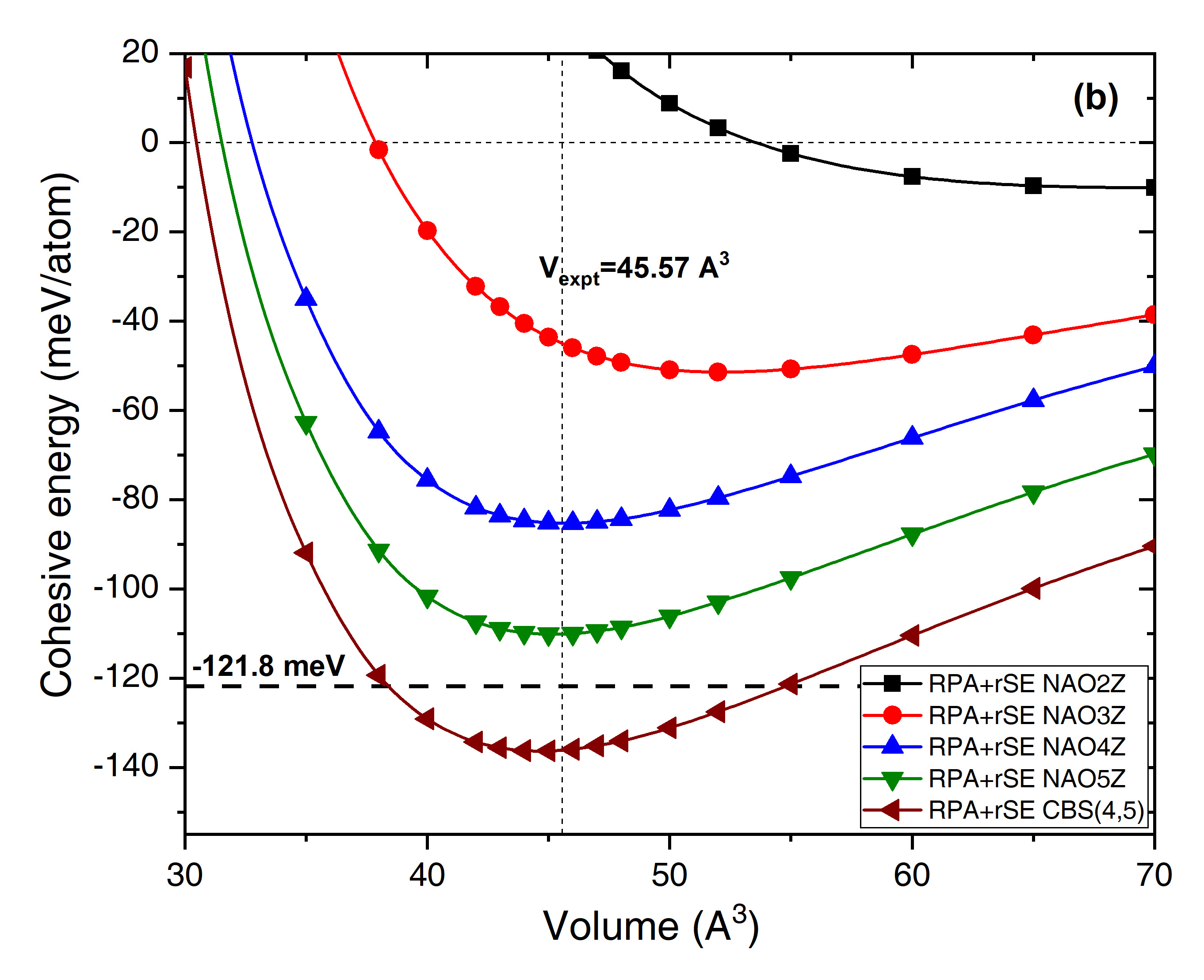}
}
\caption{\label{Fig:eq_fit_fcc_primitive} Fitting curve under second-order Birch–Murnaghan equation of state method for the FCC Kr crystal. The experimental cohesive energy corrected by zero-point vibration energy from accurate CCSD(T) is shown as black dashed line, -121.8 meV.}
\end{figure*}

\begin{figure}[t]
\centering
\includegraphics[width=0.45\textwidth,clip]{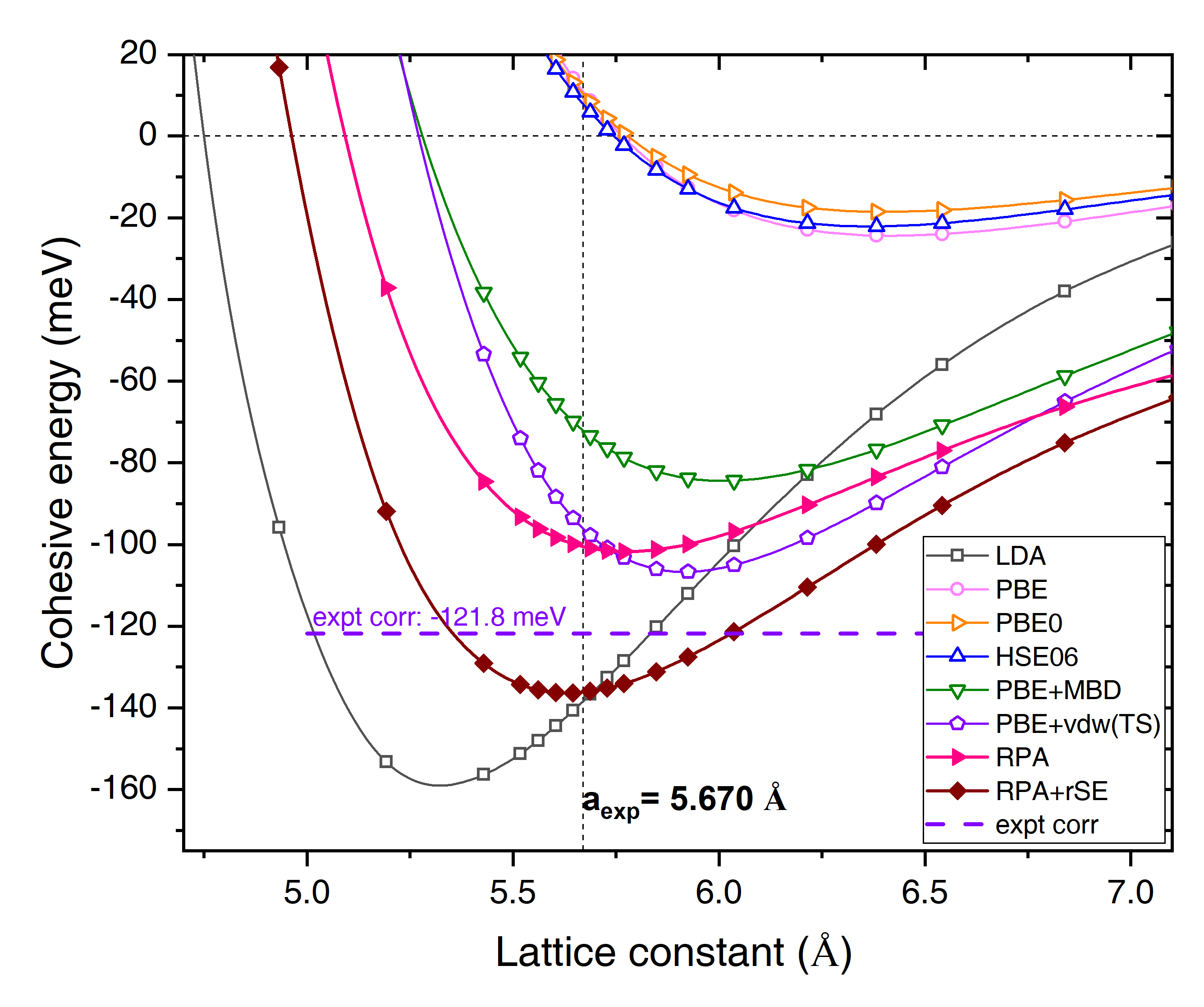}
\caption{\label{Fig:Cohesive} Cohesive energies of the FCC Kr crystal determined by different functionals. The RPA and RPA+rSE results are obtained by extrapolating the NAO-VCC-4Z and 5Z results to the complete basis set (CBS) limit, and all other results are obtained using the NAO-VCC-5Z basis set. A FCC primitive unit cell and $8\times 8\times 8$ k grid sampling are used in all calculations. Experimental values are corrected with zero-point vibration energies by accurate quantum chemical method from reference \cite{schwalbe1977thermodynamic,losee1968thermal,PhysRevB.62.5482}.}.
\end{figure}

Next, we examine the performance of NAO-VCC-$n$Z basis sets in calculating cohesive energies using RPA-type methods for bulk systems.  Figure~\ref{Fig:eq_fit_fcc_primitive} presents the cohesive energy curves for Kr FCC crystal based on RPA (left side) and RPA+rSE (right side) methods using NAO-VCC-$n$Z basis sets with $n$=2-5. As before, the figure also includes the extrapolated CBS(4,5) results. A consistent convergence pattern relative to the cardinal number $n$ is again observed.  However, it's worth noting that for bulk Kr, the results obtained using 2Z or 3Z basis sets are still
quite far from the CBS limit. Therefore, extrapolation based on larger basis sets, such as 4Z and 5Z, is necessary. Furthermore, a comparison between the left and right sides of Fig.~\ref{Fig:eq_fit_fcc_primitive} reveals that the rSE correction significantly enhances the binding strength. Concretely, RPA tends to underbind, while RPA+rSE tends to overbind the Kr crystal, respectively.

In Fig.~\ref{Fig:Cohesive}, we present the cohesive energies of the FCC Kr crystal as a function of the lattice constant calculated using RPA@PBE and (RPA+rSE)@PBE. 
For comparison, we also include results of lower-rung functionals, including LDA, GGA–PBE, the global hybrid functional PBE0 \cite{Perdew/Ernzerhof/Burke:1996}, the Heyd–Scuseria–Ernzerhof (HSE) screened hybrid functional \cite{Heyd/Scuseria/Ernzerhof:2003}, as well as PBE complemented with vdW(TS) \cite{Tkatchenko/Scheffler:2009} and with the many-body dispersion (MBD) \cite{Tkatchenko/etal:2012}. 
The experimental equilibrium lattice constant and cohesive energy are marked by dashed lines in the figure. Here the reference cohesive energy has been corrected for zero-point vibration energies as determined by accurate quantum chemical methods \cite{PhysRevB.62.5482}. Similar to other noble-gas crystals, an accurate description of the cohesive properties of the Kr crystal is  challenging for conventional DFAs. LDA and PBE exhibit substantial overbinding and underbinding tendencies, respectively. Hybrid functionals like PBE0 or HSE offer no improvement over PBE, due to their inability to capture long-range vdW interactions. Conventional GGA-PBE combined with semi-empirical vdW corrections, like PBE+vdW(TS) and PBE+MBD, improves accuracy for weak interactions but still underestimates the cohesive energy and overestimates the equilibrium lattice constant. 

Regarding the performance of RPA, we obtained an equilibrium lattice constant of $ 5.770$ \AA and a cohesive energy of -101.7 meV at the CBS(4,5) level. Compared to the experimental data of $ 5.670 \AA$ and -121.8 meV \cite{losee1968thermal}, the standard RPA@PBE scheme exhibits a noticeable underbinding, despite its significant improvement over conventional DFAs. This underbinding tendency of RPA has also been observed by Klime\v{s}~\textit{et al.} \cite{Klimes/etal:2015}, based on plane-wave-based RPA implementation \cite{Harl/Kresse:2008}. As previously discussed, RPA@PBE typically underbinds molecules and solids due to overestimated Pauli repulsion, which can be largely corrected by adding single-excitation corrections \cite{Ren/etal:2011,Ren/etal:2013,Klimes/etal:2015}. 
Figure~\ref{Fig:Cohesive} shows that incorporating the rSE correction enhances the cohesive energies of the Kr crystal, resulting in an equilibrium lattice constant of $5.633$ \AA  and a cohesive energy of -136.3 meV. While this is generally satisfactory, an overestimation 
of cohesive energy by 14 meV is noticeable, which can be partly attributed to the lack of BSSE correction in the bulk calculations.
Table.~S2 of the SM indicates that, with the CP correction, the binding energies are reduced by about 24 meV, leading to a BSSE-corrected RPA+rSE binding energy of -112.1 meV. It is worth pointing out that this value is close to the NAO-VCC-$5$Z result (see Fig~\ref{Fig:Cohesive}), indicating a 
compensation between the small BSSE and the BSIE in the finite basis set. The fully numerically converged RPA+rSE cohesive energy for Kr FCC crystal is most likely in between the CP-corrected and uncorrected values, aligning well with the experimental result.

\subsection{\label{sec:results:PV}$P$-$V$ curve of Kr FCC crystal in the high-pressure regime}
The study of matter under high pressure is an important area of research in condensed-matter physics, as a plethora of intriguing phenomena might emerge in this regime. Fascinating structural transition or abrupt changes in conductivity and other electronic-structure-derived properties are often observed in experiments. For example, it has been shown that the application of high pressure can induce a phase transition from trilayer graphene to hexagonal diamond structure \cite{ke2020synthesis}, and a structural transformation from 2H$_{c}$ to 2H$_{a}$ in MoS$_{2}$ \cite{nayak2014pressure}. At ultrahigh pressure, superconductivity has even been observed in pristine 2H$_{a}$-MoS$_{2}$  \cite{chi2018superconductivity}. Similar phenomena have been reported in the family of rare-gas solids. Specifically, a FCC-HCP phase transition has been experimentally observed \cite{jephcoat1987pressure,PhysRevLett.86.4552,goettel1989optical,PhysRevB.65.214110,Errandonea2006Structural}, accompanied by changes in the band structure \cite{PhysRevLett.62.669,PhysRevLett.95.257801}. More recently, it was found that Kr may exhibit significant stacking disorders in its FCC lattice. Furthermore, the FCC and HCP phases have been found to coexist over a wide pressure range from 2.7 to 140 GPa upon high-temperature annealing via laser heating \cite{rosa2018effect}. While considerable theoretical efforts have been devoted to the understanding of the physical properties of rare gas systems under high pressure \cite{PhysRevB.95.214116,PhysRevLett.96.035504,PhysRevB.52.15165,PhysRevLett.88.075504}, a quantitative description achieving experimental accuracy remains a significant challenge.

\begin{figure}[t]
\centering
\includegraphics[width=0.45\textwidth,clip]{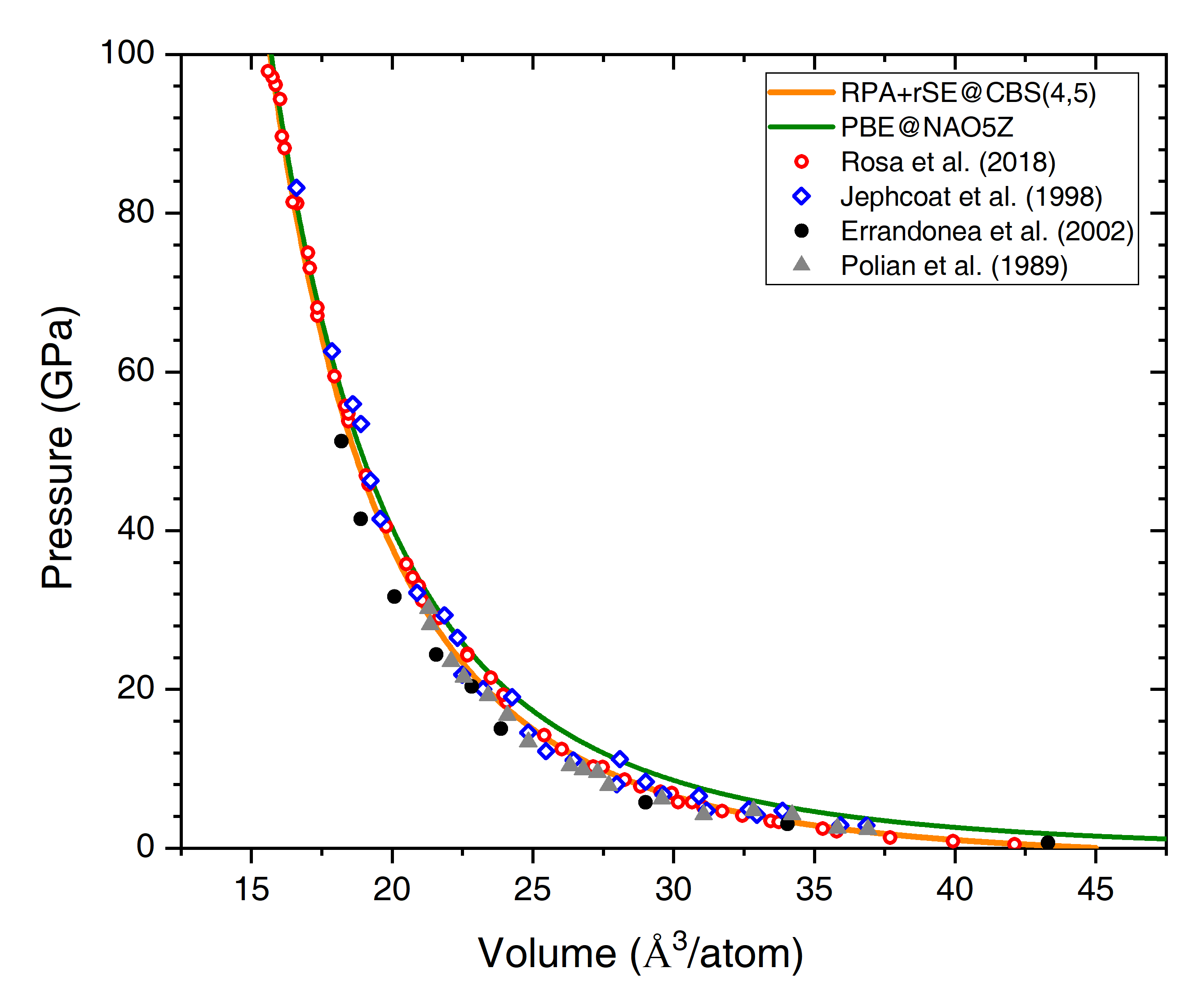}
\caption{\label{Fig:P-V diagram} Calculated $P$–$V$ diagram at 300 K for the FCC phase of the Kr crystal. The results are extrapolated to the CBS limit [CBS(4,5)]. The electronic part of the free energy is obtained using RPA+rSE, complemented with PBE phonon free energy at 300 K (orange line). Reference data from four experimental studies are included for comparison.}
\end{figure}

Having verified the reliability of our NAO-VCC-$n$Z basis set for Kr and the good performance of the RPA+rSE method in describing the cohesive property of the Kr crystal, we set out to determine the $P$-$V$ diagram theoretically. To achieve this, we calculate the Helmholtz free energy $F(V,T)$ according to Eq~\ref{Eq:QHA}. Subsequently, we determine the pressure at a given temperature and volume via $P(V,T)=-\partial F(V,T)/\partial V|_{T}$,
which allows us to generate the $P$-$V$ curves at various temperatures. In Fig.~\ref{Fig:P-V diagram}, we present the calculated (RPA+rSE)@PBE and PBE $P$-$V$ curves at $T=300$ K for the FCC phase of Kr crystal. Furthermore, we include four sets of experimental data for comparative analysis.

 
Figure~\ref{Fig:P-V diagram} shows that our calculated RPA+rSE $P$–$V$ curve is in overall good agreement with the experimental results of Polian ~\textit{et al.} \cite{polian1989x} and Jephcoat ~\textit{et al.} \cite{jephcoat1998rare} extracted from energy-dispersive x-ray diffraction data, and those of Errandonea ~\textit{et al.} \cite{PhysRevB.65.214110} and Rosa ~\textit{et al.} \cite{rosa2018effect} extracted from $\emph{in-situ}$ x-ray diffraction and absorption measurements. Note that, there are noticeable discrepancies among the various sets of
experimental data themselves. Here, we focus on the comparison to the experimental data from Rosa~\textit{et al.}, as it represents the most recent and comprehensive investigation of Kr under high-pressure conditions, extending up to 140 GPa. Overall our theoretical $P$-$V$ curve agrees well with the 
experimental data of Rosa~\textit{et al.} in both high and low pressure regimes. For comparison, we also include the calculated PBE $P$-$V$ curve in Fig.~\ref{Fig:P-V diagram}. It is evident that the PBE method tends to overestimate the pressure (i.e., a higher pressure is needed to compress the
Kr crystal to the same volume), a tendency that is pronounced in the low and medium pressure range.

\section{Conclusion}
\label{sec:conc}

In this work, we developed valence-correlation-consistent NAO basis sets for the Kr element. Our tests on the binding energies of Kr$_2$ and the cohesive energies of Kr FCC crystal demonstrate that these basis sets yield systematic convergence patterns for RPA-type correlated calculations. This allows for a reliable extrapolation to the CBS limit. More importantly, these newly generated NAO-VCC-$n$Z basis sets for Kr exhibit minimal BSSE, in stark contrast with the Gaussian cc-pVXZ basis sets and the FHI-aims-2009 NAO basis sets. This feature makes them particularly useful in the correlated calculations of extended materials, where a CP correction for BSSE is often impractical. We then calculated the Helmholtz free energy and derived the $P$-$V$ curve for the FCC phase of the Kr crystal, based on converged RPA+rSE calculations for the electronic part of the free energy. The resulting theoretical $P$-$V$ curve is in excellent agreement with the most recent experiment data. The insights and methodologies developed in this study offer valuable guidance for constructing NAO-VCC basis sets for other heavy elements. Overall, our research investigates the capability of correlation-consistent NAO basis sets for heavy elements, laying the groundwork for conducting reliable correlated calculations at the RPA level and beyond in extended materials.


\begin{acknowledgments}
    The work is supported by the National Key R \& D Program of China (Grant No.~2022YFA1403800) and the National Natural Science Foundation of China (Grant Nos.~12188101, 12134012, 21973015, and 22125301), as well as by the Max Planck Partner Group for \textit{Advanced Electronic Structure Methods}, and innovative research team of high-level local universities in Shanghai and a key laboratory program of the Education Commission of Shanghai Municipality (ZDSYS14005). 
\end{acknowledgments}

\section*{Additional Information}

\textbf{Competing interest:} The authors declare no competing financial or non-financial interests.

\section*{Data Availability}
All the input and output files of the calculations involved in this work have been uploaded to the NOMAD repository and can be found under the link https://dx.doi.org/10.17172/NOMAD/2023.09.01-1.

\bibliography{CommonBib}
\end{document}

%% file: newcommand.tex
\newcommand{\bfr}{ {\bf r}} 
\newcommand{\bfq}{ {\bf q}} 
\newcommand{\bfk}{ {\bf k}} 
